\documentclass{article}
\usepackage{fortschritte}
\usepackage{amsmath}
\usepackage{amsfonts}
\usepackage{dsfont}
\usepackage{cite}
\usepackage{epsfig}
\usepackage{epic}
\usepackage{eepic}
\usepackage{subfigure}
\usepackage{rotating}

\def\be{\begin{equation}}
\def\ee{\end{equation}}
\newcommand{\bea}{\begin{eqnarray}}
\newcommand{\eea}{\end{eqnarray}}
\newcommand{\coln}{~,}
\newcommand{\pnt}{~.}
\newcommand{\AdS}{\text{AdS}}
\newcommand{\AdSS}{\AdS_{d+1}\times \text{S}^{d'+1}}
\newcommand{\twob}{\text{II}\,\text{B}}

\newcommand{\hypergeometric}[4]{F\big(#1,#2;#3;#4\big)}
\newlength{\diameter}
    
\newlength{\neglength}
         
\makeatletter
\newcommand{\redefinelabel}[1]{
  \def\@currentlabel{#1}}
\makeatother

\def\beq{\begin{equation}}                     % 
\def\eeq{\end{equation}}                       %
\def\bea{\begin{eqnarray}}                     %         %
\def\eea{\end{eqnarray}}                       %       % 
\begin {document}                 
\def\titleline{
%%%%%%%%%%%%%%%%%%%%%%%%%%%%%%%%%%%%%%%%%%%%%%
%                                            
% Insert now the text of your title.         
% Make a linebreak in the title with         
%                                            
%            \newtitleline                   
%                                            
%%%%%%%%%%%%%%%%%%%%%%%%%%%%%%%%%%%%%%%%%%%%%%%
\begin{flushright}
\normalsize{\mdseries HU Berlin-EP-04/08}
\end{flushright}
\vspace{1cm}
Comments on the scalar propagator in $\text{AdS}\times\text{S}$ and the BMN plane
  wave 
%%%%%%%%%%%%%%%%%%%%%%%%%%%%%%%%%%%%%%%%%%%%%%%         %
}
\def\email_speaker{
{\tt
%%%%%%%%%%%%%%%%%%%%%%%%%%%%%%%%%%%%%%%%%%%%%%
%                                                  
% Insert now the e-mail address of the speaker or  
% the author that should get the electronic mail   
% of the publishing house                           
%                                                  
%%%%%%%%%%%%%%%%%%%%%%%%%%%%%%%%%%%%%%%%%%%%%%         %
 csieg@physik.hu-berlin.de                   %       %
%                                            %     %%%%%%%%%%%%%%
%                                            %       %       
%%%%%%%%%%%%%%%%%%%%%%%%%%%%%%%%%%%%%%%%%%%%%%         %
}}
\def\authors{
%                                            
%  Insert now the name (names) of the author 
%  (authors).                                
%  In the case of several authors with       
%  different addresses use labels e.g.       
%                                            
%             \1ad  , \2ad  etc.             
%                                            
%  to indicate that a given author has the   
%  address number 1 , 2 , etc.               
%  Signify the person with the above e-mail  
%  address with \sp (behind the address      
%  labels, e.g.                              
%                                            
%            \2ad\sp                         
%                                            
%  when the 2nd author                       
%  happens to be the 'speaker')                                  
%%%%%%%%%%%%%%%%%%%%%%%%%%%%%%%%%%%%%%%%%%%%%%         %
H. Dorn, M. Salizzoni, C. Sieg\1ad           %       %
                                             %     %%%%%%%%%%%%%
%                                            %       %
%%%%%%%%%%%%%%%%%%%%%%%%%%%%%%%%%%%%%%%%%%%%%%         %
}
\def\addresses{
%%%%%%%%%%%%%%%%%%%%%%%%%%%%%%%%%%%%%%%%%%%%%%
%                                            
% List now the address. In the case of       
% several addresses list them by numbers     
% using e.g.                                 
%                                            
%             \1ad , \2ad   etc.             
%                                            
% to numerate address 1 , 2 , etc.           
%                                            
%%%%%%%%%%%%%%%%%%%%%%%%%%%%%%%%%%%%%%%%%%%%%
\1ad                                        %          
Humboldt--Universit\"at zu Berlin, Institut f\"ur Physik\\
Newtonstra\ss e 15, D-12489 Berlin          %         %
                                            %       %
                                            %     %%%%%%%%%%%%%%%%%
                                            %       %
                                            %         %
                                            %          
                                            %
%%%%%%%%%%%%%%%%%%%%%%%%%%%%%%%%%%%%%%%%%%%%%
}
\def\abstracttext{
%%%%%%%%%%%%%%%%%%%%%%%%%%%%%%%%%%%%%%%%%%%%%
%                                             
% Insert now the text of your abstract.       
%                                             
%%%%%%%%%%%%%%%%%%%%%%%%%%%%%%%%%%%%%%%%%%%%%         %
We discuss the scalar propagator on generic $\AdSS$ backgrounds.
For the conformally flat situations and masses corresponding to Weyl
invariant actions the propagator is powerlike in the sum of the
chordal distances with respect to $\AdS_{d+1}$ and $\text{S}^{d'+1}$.
In all other cases the propagator depends on both chordal distances
separately. We discuss the KK mode summation to construct the propagator 
in brief.
For
$\text{AdS}_5\times\text{S}^5$ we relate our propagator to the expression
in the BMN plane wave limit and find a geometric interpretation of the 
variables occurring in the known explicit construction on the plane wave.
%%%%%%%%%%%%%%%%%%%%%%%%%%%%%%%%%%%%%%%%%%%%%         %
}
\large
\makefront
%%%%%%%%%%%%%%%%%%%%%%%%%%%%%%%%%%%%%%%%%%%%%%%%
%                                              %
%  Insert now the remaining parts of           %
%  your article.                               %
%                                              %
%%%%%%%%%%%%%%%%%%%%%%%%%%%%%%%%%%%%%%%%%%%%%%%%
\section{Introduction}
The AdS/CFT correspondence %\cite{Maldacena:1998re} 
relates $\mathcal{N}=4$
super Yang-Mills gauge theory in Min\-kow\-ski space to type  $\twob$ string
theory in $\AdS_5\times\text{S}^5$ with some RR background flux. 
%In the supergravity approximation one handles the fields in a Kaluza-Klein
%mode expansion with respect to the $\text{S}^5$. For calculations on the
%supergravity side the propagators of the whole spectrum of fields in the
%$\AdS_5$ background are an essential technical ingredient. The simplest case
%to start with is of course the well known scalar propagator
%\cite{Burgess:1985ti,D'Hoker:2002aw}. 
Explicit tests of the AdS/CFT
correspondence, beyond the supergravity approximation, remain a difficult
task, since 
the relevant string spectrum in general is not available. 
In a limit proposed by Berenstein, Maldacena and Nastase (BMN limit)
\cite{Berenstein:2002jq} 
the $\AdS_5\times\text{S}^5$ background itself is transformed via a Penrose
limit %\cite{Penrose:1976} 
to a certain plane wave spacetime
\cite{Blau:2001ne,Blau:2002dy,Blau:2002mw}. In this background string theory
is exactly quantizable%\cite{Metsaev:2001bj}
, and thus enables independent checks of the duality, including string effects. In this BMN plane wave spacetime the separation between the $\AdS_5$ and the $\text{S}^5$ part breaks down, and one has to take the limit on full 10-dimensional $\AdS_5\times\text{S}^5$ objects. 

One of the crucial unsolved questions in this setting concerns the issue
of
holography \cite{Berenstein:2002sa,Das:2002cw,Leigh:2002pt,Kiritsis:2002kz,Dobashi:2002ar}
.  
In the Penrose limiting process the old 4-dimensional conformal boundary is put beyond the new plane wave space, which by itself has
a one-dimensional conformal boundary. In \cite{Dorn:2003ct} we started to investigate this issue and found for each point remaining in the final plane wave a degeneration of the cone of boundary reaching null geodesics into
a single direction. To continue this program beyond geometric properties, we
now want to study the limiting process  for field theoretical propagators.
The scalar propagator in the BMN plane wave has been constructed in \cite{Mathur:2002ry} by a direct approach leaving the issue of its derivation via a limiting process from $\AdS_5\times\text{S}^5$ as an open problem.

Here we will discuss the construction of the scalar propagator
on $\AdSS$ spaces with radii $R_1$ and $R_2$, respectively.
Allowing for generic dimensions $d$ and $d'$ as well as generic curvature radii  $R_1$ and $R_2$ is very helpful to understand the general mechanism for the construction of the propagator. 

In section \ref{waveeqapp} we discuss the differential equation for the 
propagator and its solution in conformally flat backgrounds and for masses
corresponding to Weyl invariant actions. We will compare with the pure
$\AdS_{d+1}$ case and interpret the results from a flat space perspective.

In section \ref{modesummation} we
present the KK mode summation to construct the propagator and describe how
this leads to a theorem summing up Legendre and Gegenbauer functions. 

In section \ref{pwlimit} we will discuss the plane wave limit of $\AdS_5\times\text{S}^5$ in brief. We will explicitly show that the massless propagator on the full space indeed reduces to the expression of \cite{Mathur:2002ry}.
%%%%%%%%%%%%%%%%%%%%%%%%%%%%%%%%%%%%%%%%%%%
%%%%%%%%%%%%%%%%%%%%%%%%%%%%%%%%%%%%%%%%%%%
\section{The differential equation for the propagator and its solution}\label{waveeqapp}
The scalar propagator is defined as the solution of 
\begin{equation}\label{waveeq}
(\Box_z -M^2)G(z,z')=\frac{1}{\sqrt{-g}}\delta (z,z')\coln
\end{equation}
with suitable boundary conditions at infinity. $\Box_z$ denotes the d'Alembert operator on $\AdSS$ acting on the first argument of the propagator $G(z,z')$.
In the following we denote the coordinates referring to the $\AdS_{d+1}$ factor by
$x$ and those referring to the $\text{S}^{d'+1}$  factor by $y$, i.e. $z=(x,y)$.
$\AdS_{d+1}$ and $\text{S}^{d'+1}$ can be interpreted as embeddings respectively in $\mathds{R}^{2,d}$ and in $\mathds{R}^{d'+2}$ with the help of the constraints\footnote{More precisely, $\AdS_{d+1}$ is the universal covering of the hyperboloid in $\mathds{R}^{2,d}$.}
\begin{equation}\label{defeq}
-X_0^2-X_{d+1}^2+\sum_{i=1}^d X_i^2=-R_1^2\coln\qquad\sum_{i=1}^{d'+2}Y_i^2=R_2^2\coln
\end{equation}
where $X=X(x)$, $Y=Y(y)$ depend on the coordinates $x$ and $y$, respectively.
We define the chordal distances on both spaces to be
\begin{equation}\label{uv}
u(x,x')=(X(x)-X(x'))^2\coln\qquad v(y,y')=(Y(y)-Y(y'))^2\pnt
\end{equation}
The distances have to be computed with the corresponding flat metrics of the
embedding spaces that can be read off from  \eqref{defeq}. The chordal
distance $u$ is a unique function of $x$ and $x'$ if one restricts oneself to the hyperboloid. On the
universal covering it is continued as a periodic function.

Using the homogeneity and isotropy of both $\AdS_{d+1}$ and $\text{S}^{d'+1}$ it is clear that the propagator can depend on $z,z'$ only via the chordal distances
$u(x,x')$ and $v(y,y')$.\footnote{Strictly speaking this at first applies only if
$\AdS_{d+1}$ is restricted to the hyperboloid}.
The d'Alembert operator is a direct sum of the parts on both subspaces
$\Box_z=\Box_x+\Box_y$. $\Box_x$ and $\Box_y$ can be expressed with the
help of $u$ and $v$ and first and second derivatives w.r.t $u$ and $v$ respectively.
One can now ask for a solution of \eqref{waveeq} that only depends on the total chordal distance $u+v$. It is easy to derive that such a solution exists if and only if 
\begin{equation}\label{conditions}
R_1=R_2=R\coln\qquad M^2=\frac{d'^2-d^2}{4R^2}\pnt
\end{equation} 
One first looks for a solution of the homogeneous version of \eqref{waveeq}
and then checks its short distance singularity such that it generates the
$\delta$-function on the r.h.s. of \eqref{waveeq}. Hence after fixing the 
normalization we end up with
\begin{equation}\label{AdSSprop}
G(z,z')=\frac{\Gamma(\frac{d+d'}{2})}{4\pi^{\frac{d+d'}{2}+1}}~\frac{1}{(u+v+i\epsilon)^\frac{d+d'}{2}}\pnt
\end{equation}
Under the conditions \eqref{conditions} there is a second solution of
\eqref{waveeq}, but with the $\delta$-source shifted to another position, that depends only on $(u-v)$.
\begin{equation}\label{AdSSmirprop}
\tilde G(z,z')\propto\frac{1}{(u-v+4R^2)^\frac{d+d'}{2}}\pnt
\end{equation}
Both above given results have the same asymptotic falloff. The exponent in the
denominator is just equal to $\Delta_{+}(d,M^2)$. Where
\begin{equation}\label{conformaldim}
\Delta_{\pm}(d,m^2)=\frac{1}{2}\Big(d\pm\sqrt{d^2+4m^2R_1^2}\Big)
\end{equation}
 are the conformal dimensions of CFT fields related to the scalar fields via
 the AdS/CFT correspondence. 

The above results can be compared with the propagators in pure $\AdS_{d+1}$.
One finds simple powers of the chordal distance $u$ for the mass value 
\begin{equation}\label{AdSweylmass}
m^2=\frac{1-d^2}{4R_1^2}\pnt
\end{equation}
The two solutions are
\begin{equation}\label{AdSsuperpos}
\begin{aligned}
\frac{1}{2}(G_{\Delta_-}+G_{\Delta_+})&=\frac{\Gamma(\frac{d-1}{2})}{4\pi^{\frac{d+1}{2}}}\frac{1}{u^\frac{d-1}{2}}\coln\\
\frac{1}{2}(G_{\Delta_-}-G_{\Delta_+})&=\frac{\Gamma(\frac{d-1}{2})}{4\pi^{\frac{d+1}{2}}}\frac{1}{(u+4R^2)^\frac{d-1}{2}}\pnt
\end{aligned}
\end{equation}
Where the linear combinations on the l.h.s. refer to the expression of the
general massive scalar propagator on pure
$\AdS_{d+1}$ space \cite{Burgess:1985ti,D'Hoker:2002aw}
\begin{equation}\label{AdSprop}
G_{\Delta_\pm}(x,x')=\frac{\Gamma(\Delta_\pm)}{R_1^{d-1}2\pi^\frac{d}{2}\Gamma(\Delta_\pm-\frac{d}{2}+1)}\Big(\frac{\xi}{2}\Big)^{\Delta_\pm}\hypergeometric{\tfrac{\Delta_\pm}{2}}{\tfrac{\Delta_\pm}{2}+\tfrac{1}{2}}{\Delta_\pm-\tfrac{d}{2}+1}{\xi^2}\coln\quad\xi=\frac{2R_1^2}{u+2R_1^2}\coln
\end{equation}
evaluated for $\Delta_\pm=\frac{d\pm 1}{2}$ that corresponds to the mass value
\eqref{AdSweylmass}.  
The powerlike solution with the correct short distance behavior is given by
the first line in \eqref{AdSprop}. The second combination resembles
\eqref{AdSSmirprop}. In contrast to the $\AdSS$ case here the exponent of $u$
is given by $\Delta_-$. \\

At the end of this section we give a simple interpretation of the
conditions \eqref{conditions}. The equality of the radii is exactly the
condition for conformal flatness of the complete product space $\AdSS$ as a
whole. The mass values in
\eqref{conditions} and \eqref{AdSweylmass} are generated by the Weyl invariant
coupling of the scalar field to the background.
Therefore one can use a Weyl transformation to map
\eqref{AdSSprop} and \eqref{AdSSmirprop} and \eqref{AdSsuperpos} to flat space solutions.
The Poincar\'e patch of $\AdSS$ is mapped to $\mathds{R}^{1,d+d'+1}$ with the
boundary of $\AdSS$ mapped to a certain $d$-dimensional subspace 
$\mathds{R}^{1,d-1}$. The Poincar\'e patch 
of pure $\AdS_{d+1}$ is mapped to a flat half space $\mathds{R}_+^{1,d}$
with the boundary of $\AdS_{d+1}$ mapped to the boundary of the half space.
Therefore in the pure $\AdS_{d+1}$ case we can use the standard mirror charge
method to implement either Dirichlet or Neumann boundary conditions. These 
two solutions refer to the two values $\Delta_\pm$. 
In the $\AdSS$ case the mirror point lies inside the patch and therefore 
the mirror charge method is not applicable and the expression
\eqref{AdSSprop} is the single solution. 
This observation fits nicely with the fact that in AdS spaces one has to
respect the Breitenlohner-Freedman
bounds \cite{Breitenlohner:1982bm,Breitenlohner:1982jf}.
In the above conformally flat and Weyl invariant coupled cases the bounds 
teach us that 
two solutions (with $\Delta _+$ and  $\Delta _-$) are allowed in the pure 
$\AdS_{d+1}$ space whereas only one solution (scaling with $\Delta _+$) is
allowed in $\AdSS$. 

%%%%%%%%%%%%%%%%%%%%%%%%%%%%%%%%%%%%%%%%%%%%%%%%%
%%%%%%%%%%%%%%%%%%%%%%%%%%%%%%%%%%%%%%%%%%%%%%%%%%

\section{Mode summation on $\AdSS$}\label{modesummation}
In this section we will use the propagator on
pure $\text{AdS}_{d+1}$ \eqref{AdSprop}
and the spherical harmonics on $\text{S}^{d'+1}$ to construct the
propagator on $\AdSS$ via its mode expansion, summing up all the KK modes. 
The mode summation allows for a relaxation of the conditions
\eqref{conditions}: a conformally 
flat background ($R_1=R_2$) is no longer required but the restriction to 
a special mass value remains to ensure that the  
conformal dimensions $\Delta_\pm$ are linear functions of $l$ \footnote{This
  condition is necessary to evaluate the KK mode sum explicitly.}, with $l$
denoting the $l$th  mode in the KK tower. This is the condition to find an 
explicit expression for the sum and it is fulfilled in 
many cases.\footnote{E.\ g.\ in type $\twob$ supergravity on
$\text{AdS}_5\times\text{S}^5$ the $\Delta_\pm$ of the scalar modes
corresponding to the CPOs and descendant operators depend linearly on $l$ 
\cite{Kim:1985ez,Gunaydin:1985fk,Lee:1998bx}.} 
In the literature it is believed that an explicit computation of the KK mode
summation is too cumbersome \cite{Mathur:2002ry}. Here we will compare the
mode summation in the $\AdSS$ case for equal radii with the 
solution of the differential equation \eqref{AdSSprop}. This leads to a
theorem for summing certain products of Legendre and Gegenbauer functions.
An explicit discussion of the $\AdS_3\times\text{S}^3$ case with generic radii
and an explanation of how to deal with the mode summation can be found 
in \cite{Dorn:2003au}.

For the solution of \eqref{waveeq} we make the following ansatz
\begin{equation}\label{AdSSpropmodeexp}
G(z,z')=\frac{1}{R_2^{d'+1}}\sum_{I}G_I(x,x')Y^I(y)Y^I(y')\coln
\end{equation} 
where we sum over the multiindex $I=(l,m_1,\dots,m_{d'})$ such that $l\ge
m_1\ge\dots\ge m_{d'-1} \ge | m_{d'}|\ge 0$  and $Y^I$ denote the spherical
harmonics on $\text{S}^{d'+1}$ that are eigenfunctions with respect to the 
Laplacian on the sphere 
\begin{equation}\label{SHcasimir}
\Box_y Y^I(y)=-\frac{l(l+d')}{R_2^2} Y^I(y)\pnt
\end{equation}
The mode dependent Green's function on $\AdS_{d+1}$ then fulfills
\begin{equation}
\Big(\Box_x - m^2\Big)G_I(x,x')=\frac{1}{\sqrt{-g_\text{AdS}}}\delta
(x,x')\coln\qquad m^2=M^2+m_\text{KK}^2=M^2+\frac{l(l+d')}{R_2^2}\pnt
\end{equation} 
The solution of this equation was already given in \eqref{AdSprop}, into which
the (now KK mode dependent) conformal dimensions defined in
\eqref{conformaldim} enter. 
In the conformally flat case at the Weyl invariant mass value
\eqref{conditions} the conformal dimensions are given by $\Delta=\Delta_+=l+\frac{d+d'}{2}$.
The propagator is expressed as
\begin{equation}\label{modesumprop}
\begin{aligned}
G(z,z')&=\frac{\Gamma(\frac{d'}{2})}{4\pi}\Big(\frac{\xi}{2\pi R^2}\Big)^{\frac{d+d'}{2}}\\
&\phantom{={}}
\times\sum_{l=0}^\infty\frac{\Gamma(l+\frac{d+d'}{2})}{\Gamma(l+\frac{d'}{2})}\Big(\frac{\xi}{2}\Big)^l\hypergeometric{\tfrac{l}{2}+\tfrac{d+d'}{4}}{\tfrac{l}{2}+\tfrac{d+d'}{4}+\tfrac{1}{2}}{l+\tfrac{d'}{2}+1}{\xi^2}C_l^{(\frac{d'}{2})}(1-\tfrac{v}{2R^2})\coln
\end{aligned}
\end{equation} 
where \eqref{AdSprop} and the completeness relation for the spherical
harmonics has been used.
The evaluation of the above sum must precisely lead to the simple result
\eqref{AdSSprop}. In \cite{Dorn:2003au} we have proven a theorem that allows to
perform the above summation with integer $d$ and $d'$ as a special case.

%%%%%%%%%%%%%%%%%%%%%%%%%%%%%%%%%%%
%%%%%%%%%%%%%%%%%%%%%%%%%%%%%%%%%%%%%%
\section{The plane wave limit}\label{pwlimit}
The BMN plane wave background arises as a certain Penrose limit of $\AdS_5\times\text{S}^5$. The scalar propagator in the plane wave has been constructed in \cite{Mathur:2002ry}. In this section we study how this propagator in the massless case arises as a limit of our $\AdS_5\times\text{S}^5$ propagator \eqref{AdSSprop}.  

This approach is in the spirit of \cite{Dorn:2003ct}, where one follows the limiting process instead of taking the limit before starting any computations. One finds a simple interpretation of certain functions of the coordinates introduced in \cite{Mathur:2002ry}.

Taking the aforementioned Penrose limit of $\AdS_5\times\text{S}^5$ means to focus into the neighbourhood of a certain null geodesic which runs along an equator of the sphere with velocity of light. The metric of  $\AdS_5\times\text{S}^5$ in global coordinates
\begin{equation}
ds^2=R^2\big(-dt^2\cosh^2\rho+d\rho^2+\sinh^2\rho d\Omega_3^2+d\psi^2\cos^2\vartheta+d\vartheta^2+\sin^2\vartheta d\tilde\Omega_3^2\big)
\end{equation}
via the replacements
\begin{equation}\label{vartraf}
t=z^++\frac{z^-}{R^2}\coln\qquad\psi=z^+-\frac{z^-}{R^2}\coln\qquad\rho=\frac{r}{R}\coln\qquad\vartheta =\frac{y}{R}
\end{equation}
in the $R\to\infty$ limit turns into the BMN plane wave metric.

After expressing the chordal distances \eqref{uv} in global coordinates and 
applying \eqref{vartraf} one gets at large $R$ up to terms vanishing for $R\to\infty$
\begin{equation}\label{chordalpw}
\begin{aligned}
u&=2R^2\Big[-1+\cos\Delta z^++\frac{1}{R^2}\Big(-(\vec x^2+\vec x{\hspace{0.5pt}}'^2)\sin^2\frac{\Delta z^+}{2}+\frac{(\vec x-\vec x{\hspace{0.5pt}}')^2}{2}-\Delta z^-\sin\Delta z^+\Big)\Big]\\
v&=2R^2\Big[+1-\cos\Delta z^++\frac{1}{R^2}\Big(-(\vec y^2+\vec y{\hspace{1pt}}'^2)\sin^2\frac{\Delta z^+}{2}+\frac{(\vec y-\vec y{\hspace{1pt}}')^2}{2}-\Delta z^-\sin\Delta z^+\Big)\Big]\coln
\end{aligned}
\end{equation}
where $\Delta z^\pm=z^\pm-z'^\pm$.
In the $R\to\infty$ limit the sum of both chordal distances is thus given by
\begin{equation}\label{totalchordalpw}
\Phi=\lim_{R\to\infty}(u+v)=-2(\vec z^2+\vec z{\hspace{1pt}}'^2)\sin^2\frac{\Delta z^+}{2}+(\vec z-\vec z{\hspace{1pt}}')^2-4\Delta z^-\sin\Delta z^+\coln
\end{equation}
where $\vec z=(\vec x,\vec y)$, $\vec z{\hspace{1pt}}'=(\vec x{\hspace{0.5pt}}',\vec y{\hspace{1pt}}')$ and $\Phi$ refers
to the notation of \cite{Mathur:2002ry}. $\Phi$ is precisely the $R\to\infty$
limit of the total chordal distance on $\AdS_5\times\text{S}^5$, which remains
finite as both $\sim R^2$ terms in \eqref{chordalpw} cancel. This happens due
to the expansion around a \emph{null} geodesic.

The massless propagator in the plane wave background in the $R\to\infty$ limit of \eqref{AdSSprop} with $d=d'=4$ thus becomes
\begin{equation}
G_\text{pw}(z,z')=\frac{3}{2\pi^5}\frac{1}{(\Phi+i\epsilon)^4}\coln
\end{equation}  
which agrees with \cite{Mathur:2002ry}. 

%%%%%%%%%%%%%%%%%%%%%%%%%%%%%%%%%%%%
%%%%%%%%%%%%%%%%%%%%%%%%%%%%%%%%%%%%
\section{Conclusions}

In this paper we have focussed on the propagator of scalar
fields on $\AdSS$ backgrounds. We have discussed the defining wave equation
with $\delta$-source in this background. On conformally flat backgrounds for
Weyl invariant coupled fields the  propagator is simply powerlike in the sum
of both chordal distances. 
An interpretation from the flat space point of view was given using the mirror
charge method, resulting in an explanation why on pure AdS one finds two solutions with different asymptotic behavior whereas in $\AdSS$ there is a unique
solution. 

In addition for $\AdSS$ we have investigated the KK decomposition of the
propagator using spherical harmonics. In brief we described how a 
comparison with the solution of the differential equation lead to the 
formulation of a theorem that sums a product of a Legendre and a Gegenbauer 
function. 

For $\AdS_5\times\text{S}^5$ backgrounds we explicitly performed the Penrose limit on our expression for the propagator to find the result on the plane wave
background in agreement with \cite{Mathur:2002ry}. The coordinate dependence
is given by the $R\to\infty$ limit of the total chordal distance of 
$\AdS_5\times\text{S}^5$.

Clearly future work is necessary to construct the propagator for the case of generic mass values. But already with our results one should be able to address the issue of defining a bulk-to-boundary propagator and study its behavior in the plane wave limit.\\

%%%%%%%%%%%%%%%%%%%%%%%%%%%%%%%%%%%%
%%%%%%%%%%%%%%%%%%%%%%%%%%%%%%%%%%%%

%%%%%%%%%%%%%%%%%%%%%%%%%%%%%%%%%%%%%%%%%%% 
{\bf Acknowledgement} The work was supported by DFG (German Science Foundation) with the ``Schwerpunktprogramm Stringtheorie" and the ``Graduiertenkolleg 271".

%%%%%%%%%%%%%%%%%%%%%%
\bibliographystyle{utphys}
\bibliography{references}
%\begin{thebibliography}{77}
%
%\end{thebibliography}
\end{document}